**Resistively detected nuclear magnetic resonance via a single InSb two-dimensional electron gas at high temperature**

K.F. Yang,[1] H.W. Liu,[1, 2,a] K. Nagase,[1] T.D. Mishima,[3] M.B. Santos,[3] and Y. Hirayama[1,4,b]

[1]ERATO Nuclear Spin Electronics Project, Sendai, Miyagi 980-8578, Japan

[2]*State Key Laboratory of Superhard Materials and Institute of Atomic and Molecular Physics, Jilin University, Changchun 130012, People's Republic of China*

[3]*Homer L. Dodge Department of Physics and Astronomy, University of Oklahoma, 440 West Brooks, Norman, OK 73019-2061, USA*

[4]*Department of Physics, Tohoku University, Sendai, Miyagi 980-8578, Japan*

We report on the demonstration of the resistively detected nuclear magnetic resonance (RDNMR) of a single InSb two-dimensional electron gas (2DEG) at elevated temperatures up to 4 K. The RDNMR signal of $^{115}$In in the simplest pseudospin quantum Hall ferromagnet triggered by a large direct current shows a peak-dip line shape, where the nuclear relaxation time $T_1$ at the peak and the dip is different but almost temperature independent. The large Zeeman, cyclotron, and exchange energy scales of the InSb 2DEG contribute to the persistence of the RDNMR signal at high temperatures.

Nuclear spins in low-dimensional semiconductor structures have attracted growing interest because the long spin coherence time of nuclei facilitates the implementation of quantum information processing (QIP).[1] The contact hyperfine interaction between electrons and nuclei provides the basis for control and detection of isolated nuclear spins via the electron spin. Electrical manipulation of individual nuclei of Ga and As has been achieved in a GaAs two-dimensional electron gas (2DEG) by taking advantage of the resistively detected nuclear magnetic resonance (RDNMR), an emerging high-sensitivity NMR technique.[2-6] However, all of these experiments were performed only at millikelvin temperatures. More recently, electrical detection of coherent oscillations of nuclear spins has been realized in the breakdown regime of integer quantum Hall states of the GaAs 2DEG with an operating temperature up to 1.5 K.[7] Control and monitor of spin transfer between electrons and nuclei at elevated temperatures is undoubtedly valuable for practical applications of the nuclear spin based QIP.

We have recently developed a dynamic nuclear polarization (DNP) approach to demonstrate the RDNMR signal of the high-nuclear-spin isotopes of In and Sb in a single 2DEG within the typical narrow-gap semiconductor InSb.[8] The large effective $g$ factor ($g^*$) of the InSb 2DEG supports the formation of the simplest quantum Hall ferromagnet (QHF) featured in a resistance spike at the Landau-level (LL) intersection around the filling factor of $\nu = 2$. The domain structures in the QHF favor the DNP. The large exchange energy at the LL intersection and the large cyclotron energy due to a small effective mass ($m^*$) of the InSb 2DEG ensure the persistence of the resistance spike within the longitudinal resistance ($R_{xx}$) minima at high temperatures, a unique property that motivates the study reported here. In this letter, we present the RDNMR signal of [115]In around the $\nu = 2$ resistance spike up to 4 K and describe a possible underlying mechanism by probing the temperature dependence of the nuclear spin relaxation time ($T_1$) of [115]In.

The 2DEG in a Hall bar pattern (80 μm × 30 μm) used here is confined to a 20-nm-wide InSb quantum well with asymmetrically silicon delta-doped $Al_{0.2}In_{0.8}Sb$ barriers. This sample has the low-temperature electron mobility and density of $\mu = 13.6$ m$^2$/Vs and $n_s = 2.56 \times 10^{15}$ m$^{-2}$, respectively. RDNMR and tilted-magnetic-field measurements were performed in a dilution refrigerator with an *in-situ* rotator at temperatures between 100 mK and 6 K. A direct current (DC) was applied to trigger the nuclear polarization, and a small continuous-wave radiofrequency (RF) field (~ μT) generated by a single turn coil wound around the Hall bar [inset of Fig. 1(a)] was used to depolarize the nuclei. The longitudinal resistance $R_{xx}$ was measured by a low-noise preamplifier.

The large $g^*$ of the single InSb 2DEG[9,10] contributes to bring the lowest two Landau levels (LLs) with only orbit (n = 0,1) and spin ($\sigma = \uparrow, \downarrow$) indices into degeneracy around $\nu = 2$ using tilted magnetic fields. The quantum Hall state at this LL intersection has easy-axis pseudospin (subsuming both real spin and orbit degrees of freedom) anisotropy, showing Ising ferromagnetism.[11] Disorder or finite temperature produces a domain wall separating domains with different pseudospin polarizations $\left| (0, \downarrow), (0, \uparrow) \right\rangle$ and $\left| (0, \downarrow), (1, \downarrow) \right\rangle$.[12] Electron scattering in the domain wall leads to a resistance spike within persistent $R_{xx}$ minima as a signature of the QHF. The energetically degenerate domains with different pseudospins at the LL intersection support electron-nuclear flip-flop and thus the RDNMR demonstration in the $\nu = 2$ QHF by application of a large alternating current (AC).[8] In this work, we focus on the RDNMR study of the InSb 2DEG at elevated temperatures under a large DC bias.

By tilting the 2DEG away from the total magnetic field $B$ [$\theta$ is the tilt angle, inset of Fig. 1(a)], the ratio of Zeeman ($\propto B$) to cyclotron ($\propto B_\perp / m^*$, where $B_\perp$ is the perpendicular component of $B$) energy is tuned to make LLs with opposite spins intersect.[13] The resistance spike of the $\nu = 2$ QHF in our sample is formed at $B \sim 12.3$ T and $\theta = 64.5°$, and its temperature ($T$) dependence is shown in Fig. 1(a) with a small DC current of $I_{DC} = 10$ nA. Apparently, the spike strongly depends on $T$. Figure 1(b) shows the spike amplitude $R_{peak}$ vs. $T$, where $R_{peak}$ increases with increasing $T$. The solid line is a fit to the data [see caption to Fig. 1(b)], suggesting that a hopping transport[14] is dominant at the LL intersection. This also indicates that the charge energy scale at the spike is much larger than the thermal energy, thereby suppressing the activation process (Arrhenius dependence). The large energy gap at the LL intersection is attributed to strong exchange correlations as reflected by a giant enhancement of $g^*$.[9,10,15] Furthermore, the cyclotron energy of the InSb 2DEG is much larger than the LL broadening due to the small $m^*$, ensuring the persistence of the spike within the $R_{xx}$ minima at high temperatures. Accordingly, the InSb 2DEG with the large $g^*$ and small $m^*$ offers the chance for the high-temperature RDNMR study. Besides $T$, $R_{peak}$ is also found to increase with increasing $I_{DC}$ [Fig. 1(b)], depending not only on electron scattering in the domain wall with respect to the DNP[8] but also on heating effects. The data in Fig. 1(b) give an estimate of the upper limit of the effective temperature $T_e$ induced by $I_{DC}$ (e.g., $T_e$ should be less than 2 K at $I_{DC} = 3$ μA).

The RDNMR and $T_1$ measurements were performed following the method described in Ref. 8 but using a DC current. Figure 2 shows the RDNMR signal of [115]In at $I_{DC} = 870$ nA and $T = 100$ mK, where $\Delta R_{xx}$ is the resistance change with respect to the nuclear depolarization. In contrast to the RDNMR spectrum with only a dip or a peak in $R_{xx}$ under the AC bias,[8] the DC counterpart shows a peak-dip line shape. Such a dispersive RDNMR signal has also been demonstrated in GaAs 2DEGs,[16-20] but the underlying mechanism remains obscure. Because the DC-related RDNMR signal is stronger than the AC signal in this sample, we therefore employ the DC current for the high-temperature RDNMR study.

The absolute value of $\Delta R_{xx}$ at the peak (○) and the dip (●) of the dispersive line shape is found to decrease as $T$ increases. Examples of the RDNMR spectra of [115]In at 1 K and 4 K are shown in the upper panel of Fig. 2. Note that we apply a relatively large RF power and $I_{DC}$ to attain the distinguishable RDNMR signal at 4 K. Although the $T$-dependent $\Delta R_{xx}$ appears to be accordant with expectations, the $T$-dependent $T_1$ at both peak and dip exhibits unexpected behavior. The procedure for the $T_1$ measurement is shown in the inset of Fig. 3. Figure 3 shows $T_1$ vs. $T$ at the peak and the dip of the RDNMR signal, in which two prominent features should be addressed. The first is that $T_1$ at the peak is about six times longer than that at the dip, and the other is that $T_1$ at both positions is almost independent of $T$ up to 3 K ($T_1$ at 4 K is not attained because of the weak signal). The difference in $T_1$ at the peak and the dip of the RDNMR signal has been reported in the GaAs 2DEG, but $T_1$ increases (unusual dependence) or decreases (Korringa-law dependence) with increasing $T$ at different $\nu$.[18] The $T$-independent $T_1$ was also shown in the ferromagnetic state of the GaAs 2DEG up to 250 mK, where the disordered collective spin textures (akin to Skyrmions) were expected to dominate the nuclear relaxation process.[5,21] We here assign the Skyrmion-like spin texture in the domain wall to account for the short and temperature-independent $T_1$ in the QHF of the InSb 2DEG. Accordingly, $T_1$ does not depend on $T$ provided the collective spin textures in the domain wall persist. Note that the structural divergence of the spin textures probably results in the difference of $T_1$ at the peak and the dip of the DC signal and at the dip of the AC signal.[8] Furthermore, we suggest that the temperature at which the RDNMR signal vanishes (a little above 4 K in this study) corresponds to the Curie temperature ($T_C$) of the Ising QHF. As $T$ is higher than $T_C$, the domain wall expands to the sample perimeters,[12] destroying the magnetic order and thus the DNP. It is noteworthy that a detailed analysis of the resistance spike of the $\nu = 3$ QHF in the AlAs 2DEG gives $T_C \sim 0.5$ K.[22] Because $T_C$ is proportional to the exchange energy $E_C$, a large $E_C$ corresponds to a relatively high $T_C$. $E_C$ in the $\nu = 2$ QHF of our InSb sample is about 10 times larger than that of the $\nu = 3$ QHF of the AlAs 2DEG, the difference in $T_C$ ($\sim 10$ times) of these two samples hence looks reasonable. This interpretation also indicates that $T_C$ could be further enhanced by increasing $n_s$, because $E_C$ is large in the $\nu = 2$ QHF formed at high $B$ for large $n_s$.[9] In addition, the collapse of the domain wall at $T_C$ will affect the hopping transport via the changes in either the localized length characterized by $T_0$ [see caption to Fig. 1(b)] or in the dimensionality of the hopping electrons,[23] as reflected by deviations in the fit of the data in Fig. 1(b) around 4 K.

In summary, we have performed the RDNMR measurement on a single InSb 2DEG up to 4 K by means of a large DC current. The RDNMR study of the simplest pseudospin QHF at elevated temperatures provides unique perspectives on the dynamics of domain and DNP in the 2-D ferromagnet. The present work clearly demonstrates that the InSb 2DEG with large Zeeman, cyclotron, and exchange energies is a suitable candidate for implementation of the nuclear-spin-based information processing at high temperatures.

H.W.L. thanks the Program for New Century Excellent Talents of the University in China.

[a] Electronic email: liuhw@ncspin.jst.go.jp

[b] Electronic email: hirayama@m.tohoku.ac.jp

**FIG. 1.** (a) Temperature dependence of longitudinal resistance $R_{xx}$ vs. the total magnetic field $B$ at a tilt angle $\theta = 64.5°$ and a DC current $I_{DC} = 10$ nA. The resistance spike at the filling factor of $\nu = 2$ ($B \sim 12.3$ T) is the signature of quantum Hall ferromagnet. Inset: Setup of the tilted-magnetic-field experiment, where $\theta$ is determined from the Hall resistance. (b) Spike amplitude $R_{peak}$ in (a) as a function of $I_{DC}$ (square) at $T = 100$ mK and $T$ (circle) at $I_{DC} = 10$ nA. The solid line is a fit to the temperature dependence using the variable range hopping conductivity $\sigma_{xx} = \sigma_0 e^{(-\sqrt{T_0/T})}$ with the fitting parameters $\sigma_0 = 0.0112$ S and $T_0 = 1.89$ K. $\sigma_{xx}$ is related to the longitudinal (Hall) resistivity $\rho_{xx}$ ($\rho_{xy}$) by $\sigma_{xx} = \rho_{xx}/(\rho_{xx}^2 + \rho_{xy}^2)$ with $\rho_{xy} \approx 13$ k$\Omega$ at $\nu = 2$ and to $R_{peak}$ by $R_{peak} = \rho_{xx} \, l \, / \, w$ with length $l = 80$ μm and width $w = 30$ μm of our Hall bar.

**FIG. 2.** RDNMR signal of $^{115}$In at $T = 100$ mK with a slow RF sweep rate of 100 Hz/s and a small RF power of $P = -6$ dbm. The data was obtained by the following way: As the spike resistance $R_{xx}$ becomes saturated after a large current $I_{DC} = 870$ nA is applied at $B = 11.6$ T in Fig. 1(a), the RF field sets to irradiate the sample while sweeping the frequency ($f$) through the Larmor resonance of $^{115}$In. $\Delta R_{xx}$ is the resistance change induced by the nuclear depolarization (see Ref. 8). The fine structures in the spectrum are the resolved quadrupole splittings of $^{115}$In. Upper panel: RDNMR signals of $^{115}$In at $T = 1$ K ($I_{DC} = 870$ nA and $P = -6$ dbm) and $T = 4$ K ($I_{DC} = 1.5$ μA and $P = 2$ dbm) with a fast sweep rate of 4 kHz/s.

**FIG. 3.** (color online) Temperature dependence of the nuclear spin relaxation time $T_1$ at the peak (○) and the dip (●) of the RDNMR signal in Fig. 2. Note that large $I_{DC}$ and $P$ are applied to obtain the RDNMR signal at high $T$ ($I_{DC} = 870$ nA and $P = -6$ dbm at 100 mK and 1 K; $I_{DC} = 1.5$ μA and $P = 0$ dbm at 2 K and 3 K). Inset: Plot of $\Delta R_{xx}$ vs. time ($t$) for the $T_1$ measurement. The region between the dash lines corresponds to the "on resonance" state (short for "ON") where $f$ matches the frequency at the dip of the RDNMR signal. Beyond this region $f$ shifts to the off-resonance condition ("OFF"). $T_1$ is obtained from an exponential fit to the recovery curve of $\Delta R_{xx}$ (solid line) and is found to be independent of the stimulus condition. A similar procedure is used to measure $T_1$ at the peak of the line shape.

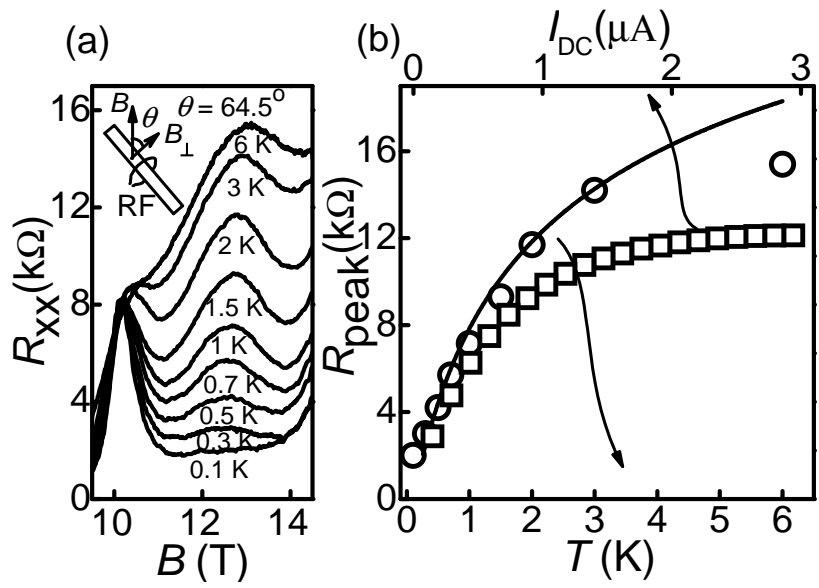

FIG. 1

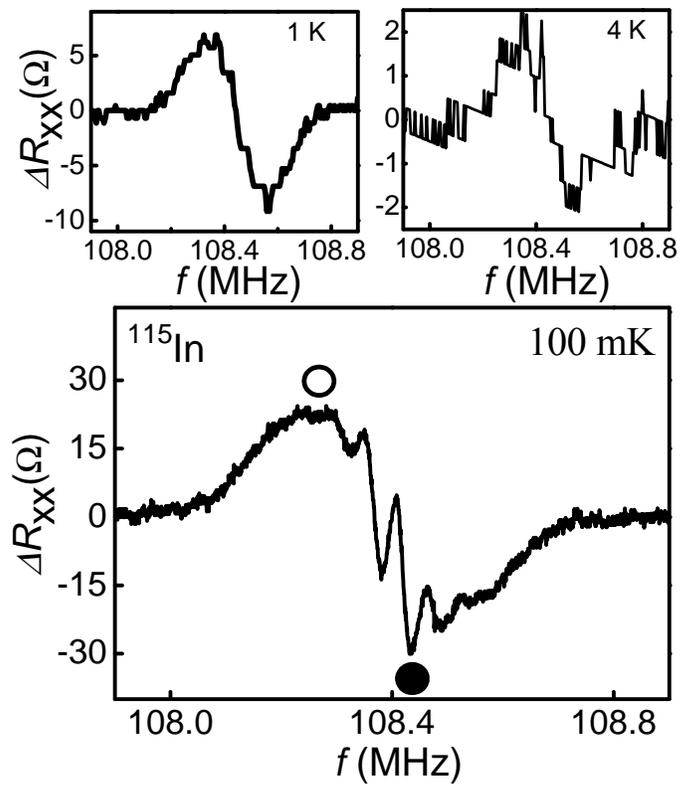

FIG. 2

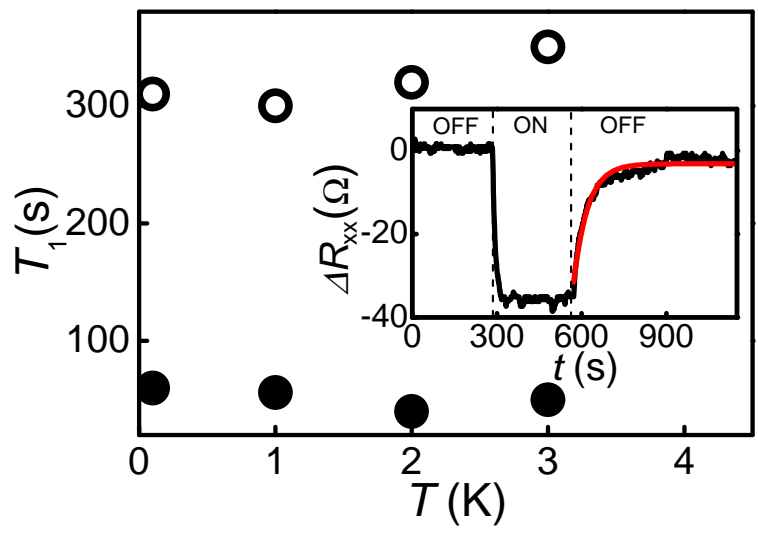

FIG. 3